# Accurate evaluation of the fractal dimension based on a single morphological image

Feng Feng[1], Binbin Liu[1], Xiangsong Zhang[1], Xiang Qian[1]*, Xinghui Li[1]*, Timing Qu[2] and Pingfa Feng[1]
1. *Division of Advanced Manufacturing, Graduate School at Shenzhen, Tsinghua University, Shenzhen 518055, China.*
2. *The State Key Laboratory of Tribology, Department of Mechanical Engineering, Tsinghua University, Beijing 100084, China.*

E-mail: qian.xiang@sz.tsinghua.edu.cn; li.xinghui@sz.tsinghua.edu.cn

**Abstract**

Fractal dimension ($D$) is an effective parameter to represent the irregularity and fragmental property of a self-affine surface, which is common in physical vapor deposited thin films. $D$ could be evaluated through the scaling performance of surface roughness by using atomic force microscopy (AFM) measurements, but lots of AFM images with different scales ($L$) are needed. In this study, a surface roughness prediction (SRP) method was proposed to evaluate $D$ values of a single AFM image, in which the roughness at smaller $L$ was estimated by image segmentation with flatten modification. Firstly, a series of artificial fractal surfaces with ideal dimension ($D_i$) values ranging from 2.1 to 2.9 were generated through Weierstrass-Mandelbrot (W-M) function, in order to compare SRP method with traditional methods such as box counting method and power spectral density method. The calculated dimension ($D_c$) by SRP method was much closer to $D_i$ than the other methods, with a mean relative error of only 0.64%. Secondly, SRP method was utilized to deal with real surfaces, which were AFM images of amorphous alumina thin films with $L$ of 1-70 μm. $D_c$ obtained by SRP method based on a single AFM image was also close to the result in our previous study by multi-image analysis at $L$ above 10 μm, while the larger $D_c$ at smaller $L$ was consisted with the actual surface feature. The validity of SRP method and the physics nature of real surfaces were discussed, which might be helpful to obtain more understandings of fractal geometry.

Surface morphology is a crucial part of thin film research to investigate the performance of functional films [1] and the deposition mechanism [2]. The morphological images of a surface could be practically measured by atomic force microscopy (AFM) [3] or other apparatus. The characteristics of surface morphology, such as roughness, are usually scale-dependent [4, 5], and fractal geometry has been widely used to analyze the scaling performances [6]. Thin films fabricated via physical vapor deposition (PVD) are generally with fractal surfaces (the self-affine type) due to the correlation of a newly absorbed atom and adjacent positions during the diffusion process [7, 8]. The fractal dimension ($D$) ranging from 2 to 3 is useful to reveal the irregularity of the surface morphology, and a larger $D$ indicates a more irregular and fragmenting surface.

There are several traditional methods commonly utilized in the literature to calculate $D$ base on the morphological image with limited resolution, such as box counting (BC) method [9, 10], power spectral density (PSD) method [11], autocorrelation function (ACF) method [12, 13], structure function (SF) method [14] and roughness scaling method [15]. Most of the above methods (ACF, SF, BC and PSD) could be applied for a single image. Generally, roughness scaling method needs a series of images with various scales ($L$) to evaluate $D$ (named as multi-image analysis method in this study), which is considered to be more accurate than the other methods according to a comparative study by Kulesza et al. [16]. The multi-image analysis method was based on a power-law relationship between root-mean-squared ($R_q$) roughness value and $L$ ($R_q \propto L^{3-D}$).

Therefore, the $R_q$-$L$ curve is linear in the double logarithmic coordinate, and $D$ can be obtained by data-fitting. Such a method was also used in our previous study [17], in which more than 20 AFM images with at least 5 different $L$ were measured for one sample.

To shorten the time consumed by measurements and calculations, the roughness scaling behavior based on an AFM single image was used in some publications [18] by segmenting the image and calculating the roughness of sub-images. However, image segmentation might bring some influence because the resolution of sub-images would be lower than that of an actually measured image at the same $L$. In our previous study [17], it was found that if a single AFM image was segmented, the roughness scaling performance of the sub-images would deviated significantly from actual AFM measurement at the same $L$. To alleviate the deviation, an additional flatten modification should be applied for each sub-image, thus the reliable roughness values could be correctly predicted in a certain $L$ range. The flatten modification was processed as following: a polynomial with a specific order (usually 2) was obtained for each scanline by least-squares fitting, then the polynomial was subtracted from the scan line to provide the final image. In this study, the roughness scaling performance obtained by segmenting a single AFM image, which could be named as surface roughness prediction (SRP) method, was used to evaluate $D$, and the validity of the SRP method was investigated and discussed to obtain accurate $D$ values.

$$z(x,y) = L\left(\frac{G}{L}\right)^{D-2}\left(\frac{\ln\gamma}{M}\right)^{1/2}$$

$$\sum_{m=1}^{M}\sum_{n=0}^{n_{max}}\gamma^{(D-3)n}\left\{\cos(\Phi_{mn}) - \cos\left[\frac{2\pi\gamma^n(x^2+y^2)^{1/2}}{L}\cos\left(\tan^{-1}\left(\frac{y}{x}\right)-\frac{\pi m}{M}\right)+\Phi_{mn}\right]\right\}$$

Two types of surfaces would be analyzed in the following. First, a series of artificial fractal surfaces with ideal dimension ($D_i$) values ranging from 2.1 to 2.9 were generated through Weierstrass-Mandelbrot (W-M) function as shown above [16]. $L$ of the artificial images was 70 μm. 25 images were generated for each $D_i$. Second, real surfaces of amorphous alumina thin films deposited by ion beam sputtering on the electro polished Hastelloy substrates were also analyzed. More details of deposition parameters and AFM measurements details could be found in our previous publication [17]. The images of both types were of 256×256 pixels, and Matlab was used for data processing.

The artificial images were firstly evaluated by using the traditional methods, as shown in Fig. 1. For the BC method, there was a significant deviation of $D_c$ when $D_i$ was large, while the deviation for ACF and SF methods occurred when $D_i$ was small. $D_c$ of PSD method was relatively accurate only when $D_i$ was in a mediate range (2.5~2.7). Therefore, the traditional methods could not provide accurate evaluation within the whole $D_i$ range, which was similar to the conclusion of Kulesza et al. [16].

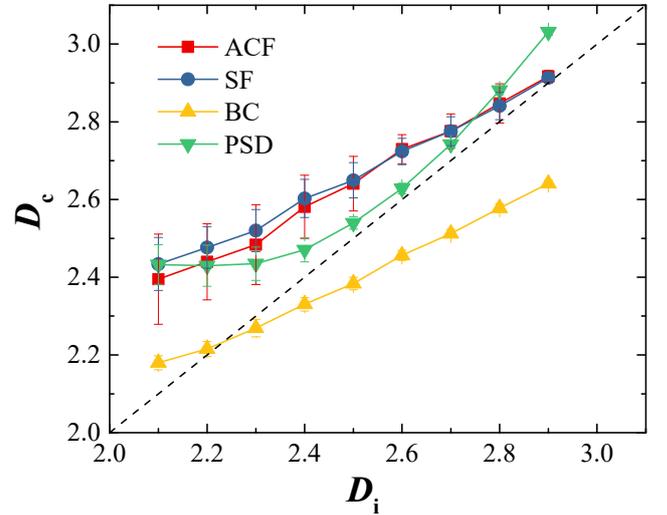

FIG. 1 The calculated fractal dimensions ($D_c$) of artificial surfaces with different ideal dimensions ($D_i$) by using autocorrelation function (ACF) method, structure function (SF) method, box counting (BC) method, and power spectral density (PSD) method.

The SRP method was then applied for each artificial image. In the image segmentation shown in Fig. 2 (a), $L$ of the sub-image became 3/4 of the last step approximately (not exactly because the pixels were integer), and the center distance of sub-images was 1/3 $L$. The segmentation stopped when the sub-image contained 4 pixels. The segmented sub-images could be utilized to calculate the roughness directly (noted as f0), or be flattened with order 1 or 2 prior to the roughness calculation (noted as f1 or f2). The roughness scaling curves obtained by SRP method could be observed in Fig. 2 (b), in which the linear portion in double logarithmic coordinate could be fitted to obtain $D_c$. It could be noticed that the first three data-points with large $L$ of f0 curve deviated from the linear portion, thus they were not fitted. On the contrary, in the f1 and f2 curves, the data-point with the smallest $L$ of deviated slightly from the linear portions.

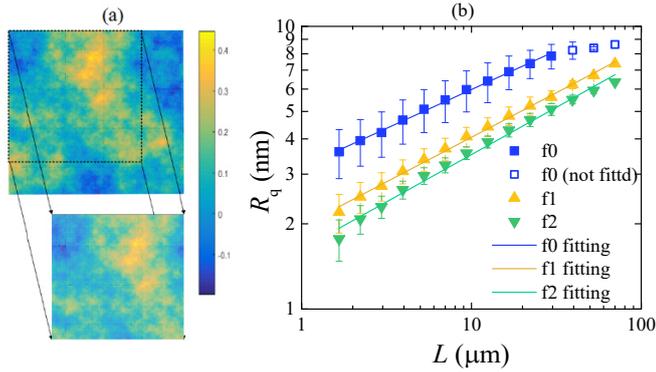

FIG. 2 The procedures in SRP method: (a) an image segmentation, (b) the roughness ($R_q$) scaling curves and fittings obtained without flatten (f0), and with flatten order of 1 or 2 (f1, f2). It could be noted that three data points at large $L$ of the f0 curve (hollow symbols) could not be fitted. $D_i$ of the artificial surface was 2.70.

$D_c$ values obtained with SRP method through fitting roughness scaling curves were summarized in Fig. 3. The result of SRP-f1 and SRP-f2 were both very close to $D_i$. $D_c$ of SRP-f0 was not accurate when $D_i$ was small, which might be due to the omitting of three data-points with large $L$ in the fitting. In order to compare all the methods, the mean relative error $<|(D_c-D_i)/D_i|>$ were listed in Table 1. For the traditional methods (ACF, SF, BC and PSD), the mean relative error ranged in around 4-7%, while those of SRP were much lower. SRP-f1 with the mean relative error of only 0.64% could be regarded as an accurate method to evaluate artificial images in the research scope of this study.

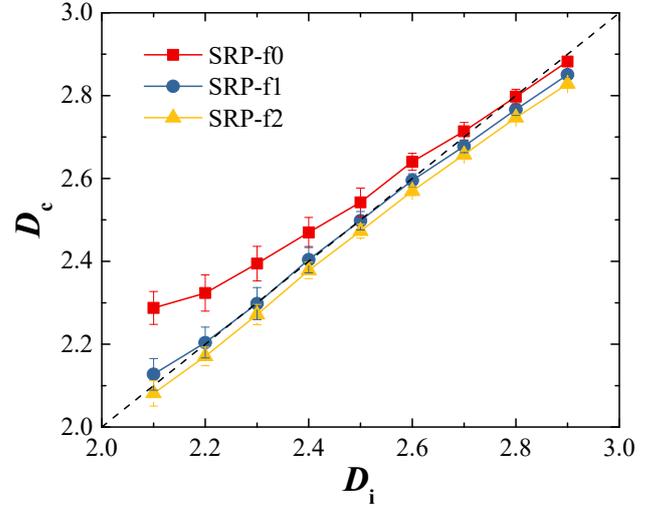

FIG. 3 $D_c$ of artificial surfaces with different $D_i$ by using SRP method without flatten (f0) and with flatten order of 1 or 2 (f1, f2).

Table 1. The mean relative error between the calculated dimension ($D_c$) and the ideal dimension ($D_i$) using different evaluation methods.

| Method | ACF | SF | BC | PSD | SRP-f0 | SRP-f1 | SRP-f2 |
|---|---|---|---|---|---|---|---|
| Mean relative error (%) | 6.24 | 6.88 | 4.75 | 5.20 | 2.89 | 0.64 | 1.40 |

In order to further verify the validity of SRP method, real surfaces of amorphous alumina thin films were also evaluated. $L$ of the AFM images were 1 μm, 3 μm, 10 μm, 20 μm, 40 μm and 70 μm, respectively. $D_c$ values obtained with all the methods were shown in Fig. 4, in which the results of BC, SRP-f1 and SRP-f2 in the $L$ range of 10-70 μm were similar to that (2.035±0.029) obtained by multi-image analysis in publication [17]. In the same $L$ range, $D_c$ of other methods were all much larger, consistent with the results of Figs. 1 and 2.

Moreover, it could be noticed that $D_c$ obtained with SRP-f1 method became larger than the result of multi-image analysis at $L$ of 1 μm and 3 μm. Such a fact could be attributed to the morphology of fine clusters and voids which could be recognized in AFM images with $L$ smaller than 10 μm, while the main surface feature was quite smooth and compact at larger $L$, as shown in Fig. (b) and (c), respectively. However, such the feature difference was not detected by using the multi-image analysis, because only one value of $D$ could be obtained using all the AFM images within a certain range of $L$. The scale-dependence of $D$ value might be common for the real surfaces since many morphological features are of a certain scale range, such as crystalline grains [19] or clusters observed in this study.

Therefore, $D_c$ result of SRP-f1 was reliable for the real surfaces in this study. SRP-f1 method could significantly shorten the measurement duration because of the single image requirement, while more than 10 images are generally required for the multi-image analysis method.

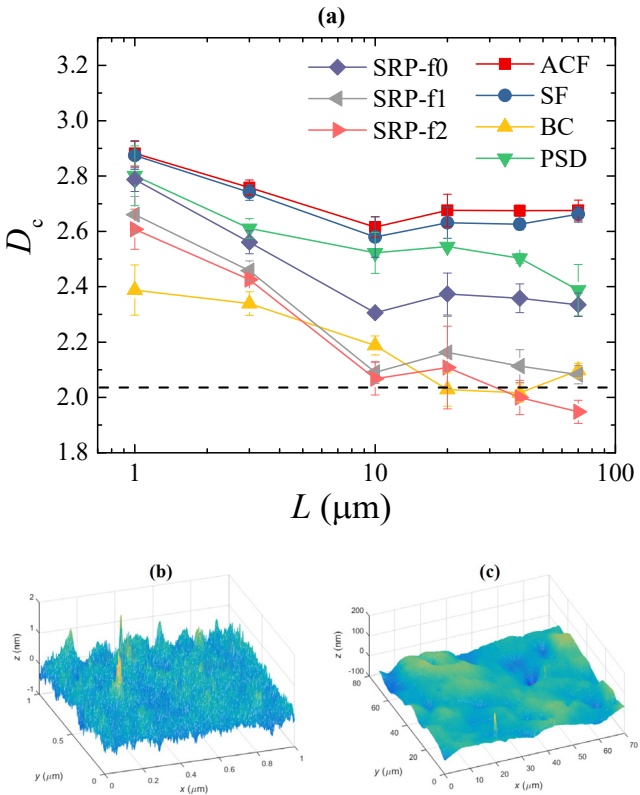

FIG. 4 (a) $D_c$ comparison of actual AFM images with different $L$ by using the methods of SRP-f0, SRP-f1, SRP-f2, ACF, SF, BC and PSD. $D$ value (2.035±0.029) obtained by roughness scaling using multi-images in our previous study [17] was marked by a dash line; (b) AFM image with $L$ of 1 μm; (c) AFM image with $L$ of 70 μm.

The validity of SRP method could be mainly attributed to the effect of flatten modification. A sub-image contains all the high-frequency components of the original image under uniform and isotropic assumption, while the low-frequency components become an uneven background. The fractal property such as roughness scaling might mainly depends on the high-frequency components, and the low-frequency components should be erased by flatten modification. However, more work should be carried out in our future study to further investigate the SRP method and make more understandings of the physics nature of fractal surfaces. During the generation of an artificial image, the truncation of the frequency range might lead to some distortion of the ideal surface from the digital data, which would be studied by changing the function parameter or using a fractal surface generating function other than the W-M one. Notwithstanding, SRP method might also be useful for other fields such as geomorphology, in which fractal geometry was commonly used.

In summary, the SRP method was proposed to obtain accurate fractal dimension based on a single morphological image. A series of artificially ideal fractal images were generated through W-M function, and different methods including ACF, SF, BC, PSD and SRP were utilized. $D_c$ result of SRP was much closer to $D_i$ than the traditional methods, because the mean relative error of SRP-f1 was only 0.64% while those of other methods were generally 4-7%. Real surfaces of amorphous alumina thin films were also applied to verify the validity of SRP method, and the scale-dependence of fractal dimension could be obtained. SRP method (particularly SRP-f1) was effective to accurately evaluate $D$ values of both artificial and real surfaces.

This study was supported by National Natural Science Foundation of China (Grant No. 51475257), Fundamental Research Program of Shenzhen (Grant No. JCYJ20170307152319957), and Tribology Science Fund of State Key Laboratory of Tribology, China. Binbin Liu and Xiangsong Zhang contributed equally to this work.